\documentclass[12pt,journal,onecolumn]{IEEEtran}

\usepackage[dvips]{graphicx}
\usepackage{epsfig}
\usepackage[cmex10]{amsmath}
\usepackage{amssymb}
\usepackage{amsthm}
\usepackage{amsfonts}
\usepackage{bm}
\usepackage{dsfont}
\usepackage[mathscr]{eucal}
\usepackage{cite}
\usepackage{pstricks}
\usepackage[tight,footnotesize]{subfigure}
\usepackage{url}
\newcommand{\remove}[1]{}

\usepackage{ifthen,epic,psfrag}
\usepackage{enumerate}

\graphicspath{{figs/}}

\input{niesen.def}

\begin{document}

\title{Computation over Mismatched Channels%
\thanks{The work of N. Karamchandani and S. Diggavi was supported in part by AFOSR MURI  award FA9550-09-064: "Information Dynamics as Foundation for Network Management". The work of U. Niesen was supported in part by AFOSR under grant FA9550-09-1-0317. The material in this paper was presented in part at the 50th Annual Allerton Conference on Communication, Control, and Computing, Monticello, IL, USA, October 2012.\newline
\indent Nikhil Karamchandani is with the Department of Electrical Engineering at the University of California Los Angeles, Los Angeles, CA 90095, USA, and the Information Theory and Applications Center at the University of California San Diego, La Jolla, CA 92093, USA (email: nikhil@ee.ucla.edu).\newline
\indent Urs Niesen is with Bell Labs, Alcatel-Lucent, Murray Hill, NJ 07974, USA 
(email: urs.niesen@alcatel-lucent.com).\newline
\indent Suhas Diggavi is with the Department of Electrical Engineering at the University of California Los Angeles, Los Angeles, CA 90095, USA (email: suhasdiggavi@ucla.edu).}
}%

\author{
Nikhil Karamchandani, Urs Niesen, and Suhas~Diggavi }

\maketitle

\begin{abstract}
    We consider the problem of distributed computation of a target
    function over a two-user deterministic multiple-access channel. If the target and channel
    functions are matched (i.e., compute the same function),
    significant performance gains can be obtained by jointly designing
    the communication and computation tasks. However, in most situations
    there is mismatch between these two functions.  In this work,
    we analyze the impact of this mismatch on the performance gains
    achievable with joint communication and computation designs over
    separation-based designs. We show that for most pairs
    of target and channel functions there is no such gain, and
    separation of communication and computation is optimal. 
\end{abstract}

\section{Introduction}
\label{sec:introduction}

The problem of computing a function from distributed information arises
in many different contexts ranging from auctions and financial trading
to sensor networks. In order to compute the desired target function,
communication between the distributed users is required.  If this
communication takes place over a shared medium, such as in a wireless
setting, the channel introduces interactions between the transmitted
signals. This suggests the possibility to harness these signal
interactions to facilitate the task of computing the desired target
function. A fundamental question is therefore whether by jointly
designing encoders and decoders for communication and computation, we
can improve the efficiency of distributed computation.

\subsection{Summary of Results}

In this paper, we explore this question by considering computation of a
function over a two-user multiple-access channel (MAC). In order to
focus on the impact of the structural mismatch between the target and
channel functions on the efficiency of computation, we ignore channel
noise and consider  only  \emph{deterministic} MACs here.  More
formally, the setting consists of two transmitters observing a (random)
variable $\msf{u}_1\in\mc{U}$ and $\msf{u}_2\in\mc{U}$, respectively,
and a receiver aiming to compute the function
$a(\msf{u}_1,\msf{u}_2)\in\mc{W}$ of these variables. The two
transmitters are connected to the destination through a deterministic
MAC with  inputs $x_1,x_2\in\mc{X}$ and output
$\msf{y}=g(x_1,x_2)\in\mc{Y}$, where $g(\cdot,\cdot)$ describes the
actions of the channel.  

A straightforward achievable scheme for this problem is to separate the
tasks of communication and computation: the transmitters communicate the
values of $\msf{u}_1$ and $\msf{u}_2$ to the destination, which then
uses these values to compute the desired target function
$a(\msf{u}_1,\msf{u}_2)$. This requires the receiver to decode
$2\log\card{\mc{U}}$ message bits. However, the MAC itself also computes
a function $g(x_1,x_2)$ of the two inputs $x_1,x_2$, creating the
opportunity of taking advantage of the structure of $g(\cdot,\cdot)$ to
calculate $a(\cdot,\cdot)$. This is trivially possible when
$g(\cdot,\cdot)$ and $a(\cdot,\cdot)$ are \emph{matched}, i.e., compute
the same function on their inputs. In such cases, performing the tasks
of communication and computation jointly results in significantly fewer
bits to be communicated. Indeed, in the matched case only the
$\log\card{\mc{W}}$ bits describing the function value are recovered at
the receiver. This could be considerably less than the
$2\log\card{\mc{U}}$ bits resulting from the separation approach.
Naturally, in most cases the channel $g(\cdot,\cdot)$ and the target
function $a(\cdot,\cdot)$ are \emph{mismatched}. The question is thus
whether we can still obtain performance gains over separation in this
mismatched situation. In other words, we ask if in general the natural
computation done by the channel can be harnessed to help with the
computation of the desired target function.

We consider two cases: i) \emph{One-shot communication}, where the MAC
is used only once, but the channel input alphabet $\mc{X}$ and output
alphabet $\mc{Y}$ are allowed to vary as a function of the domain
$\mc{U}$ of the target function. In this case, performance is measured
in terms of the scaling needed for the channel alphabets with respect to
the computation alphabets, i.e., how $\card{\mc{X}},\card{\mc{Y}}$ grow
with $\card{\mc{U}}$. This is closer to the formulation in the computer
science literature. ii) \emph{Multi-shot communication}, where the
channel alphabets $\card{\mc{X}}, \card{\mc{Y}}$ are of fixed size, but
the channel can be used several times. In this case, performance is
measured in terms of computation rate, i.e., how many channel uses are
needed to compute the target function. This is closer to the formulation
considered in information theory.

As the main result of this paper, we show that  separation between
computation and communication is essentially optimal for
most\footnote{More precisely, among all target functions
$a(\cdot,\cdot)$ with given domain $\mc{U}$ and range $\mc{W}$, and all
channel functions $g(\cdot, \cdot)$ with given input alphabet $\mc{X}$
and output alphabet $\mc{Y}$, separation is optimal except for at most
an exponentially small (in domain size $\card{\mc{U}}$) fraction of
pairs.} pairs $(a,g)$ of target and channel functions. In other words,
the structural mismatch between the functions $a(\cdot,\cdot)$ and
$g(\cdot,\cdot)$ is in general too strong for joint computation and
communication designs to yield any performance gains.

We illustrate this with an example for one-shot communication.  Assume
that the variables $\msf{u}_1,\msf{u}_2$ at the transmitters take on a
large range of values, say $\card{\mc{U}}=2^{1000}$, and the receiver is
only interested in knowing if $\msf{u}_1\geq \msf{u}_2$, i.e., in a
binary target function. Then for most MACs and one-shot communication, a
consequence\footnote{While the theorems only present results in the
limit as $\card{\mc{U}} \rightarrow \infty$, it follows from the proofs
that for a given domain $\mc{U}$ the statements hold for all but an
exponentially small (in $\card{\mc{U}}$) fraction of channel functions.}
of Theorems~\ref{Thm:IdentityOneUse} and ~\ref{Thm:balanced} in Section
\ref{sec:main} (illustrated in Example~\ref{eg:greater}) is that the
transmitters need to convey the \emph{entire} values of
$\msf{u}_1,\msf{u}_2$ to the destination, which then simply compares
them.  Thus, even though the destination is interested in only a
\emph{single} bit about $(\msf{u}_1,\msf{u}_2)$, it is still necessary
to transmit $2\log\card{\mc{U}}=2000$ bits over the channel. 

More generally, Theorems \ref{Thm:IdentityOneUse} and \ref{Thm:balanced}
in Section~\ref{sec:main} together demonstrate that for most target
functions separation of communication and computation is asymptotically
optimal for most MACs. Example~\ref{eg:equality} illustrates that only
for special functions like an equality check (i.e., checking whether
$\msf{u}_1=\msf{u}_2$) can we significantly improve upon the simple
separation scheme. Intuitively, this is because the structural mismatch
between most target and channel functions is too large to allow for any
possibility of direct computation of the target function value without
resorting to recovering the user messages first. The technical ideas
that enable these observations are based on a connection with results in
extremal graph theory such as existence of complete subgraphs and
matchings of a given size in a bipartite graph.  These connections might
be of independent interest. 

Similarly, for multi-shot communication, where we repeatedly use a fixed
channel, Theorem~\ref{Thm:converse_n} in Section~\ref{sec:main} shows
that for most functions, the computation rate is necessarily as small as
that for the identity target function describing the entire variables
$\msf{u}_1,\msf{u}_2$ at the destination. In other words, separation of
communication and computation is again optimal for most target and
channel functions. To prove this result, the usual approach using
cut-set bound arguments is not tight enough. Indeed, Example~\ref{eg:n}
shows that the ratio between the upper bound on the computation rate
obtained from the cut-set bound and the correct scaling derived in
Theorem~\ref{Thm:converse_n} can be unbounded. Rather, the structures of
the target and channel functions have to be analyzed jointly.

These results show that, in general, there is little or no benefit in
joint designs: computation-communica\-tion separation is optimal for most
cases. We thus advocate in this paper that separation of computation and
communication for multiple-access channels is not just an attractive
option from an implementation point of view, but, except for special
cases, actually entails little loss in efficiency.

\subsection{Related Work} 

The problem of distributed function computation has a rich history and
has been studied  in many different contexts. In computer science, it
has been studied under the branch of communication complexity, for
example see \cite{kushilevitz06} and references therein. Early seminal
work by Yao \cite{yao79} considered interactive communication between
two parties. Among several other important results, the paper showed
that the number of exchanged bits required to compute most target
functions is as large as for the identity function. In the context of
information theory, distributed function computation has been studied as
an extension of distributed source coding in \cite{korner79, orlitsky01,
doshi10}. For example, K\"{o}rner and Marton \cite{korner79} showed that
for the computation of the finite-field sum of correlated sources linear
codes can outperform random codes. This was extended to large networks
represented as graphs in \cite{giridhar05, appuswamy11, Nikhil11} and
references therein.  Randomized gossip algorithms \cite{Boyd06} have
been proposed as practical schemes for information dissemination in
large unreliable networks and were studied in the context of distributed
computation in \cite{Kempe03, Boyd06} among several others.

In most of these works, communication channels are represented as
orthogonal point-to-point links. When the channel itself introduces
signal interaction, as is the case for a MAC, there can be a benefit
from jointly handling the communication and computation tasks as
illustrated in \cite{cover80}. Function computation over MACs has been
studied in \cite{zhang06,katti07,nazer07,wilson10} and references
therein. 

There is some work touching on the aspect of structural mismatch between
the target and the channel functions. In \cite{niesen11}, an example was
given in which the mismatch between a linear target function with integer
coefficients and a linear channel function with real coefficients can
significantly reduce efficiency. In \cite{wilson10}, it was conjectured
that, for computation of finite-field addition over a real-addition
channel, there could be a gap between the cut-set bound and the
computation rate. In \cite{keller10}, mismatched computation when the
network performs linear finite-field operations was studied. To the best
of our knowledge, a systematic study of channel and computation mismatch
is initiated in this work.

\subsection{Organization}

The paper is organized as follows. In Section~\ref{sec:problem}, we
formally introduce the questions studied in this paper. We present the
main results along with illustrative examples in Section~\ref{sec:main}.
Most of the proofs are given in Section~\ref{sec:proofs}.

\section{Problem Setting and Notation}
\label{sec:problem}

Throughout this paper, we use sans-serif font for random variables,
e.g., $\msf{u}$. We use bold font lower and upper case to denote vectors
and matrices, e.g., $\bm{y}$ and $\bm{G}$. All sets are typeset in
calligraphic font, e.g., $\mc{X}$. We denote by $\log(\cdot)$ and
$\ln(\cdot)$ the logarithms to the base $2$ and $e$, respectively.

\begin{figure}[ht]
    \begin{center}
        \hspace{-0.5cm}\input{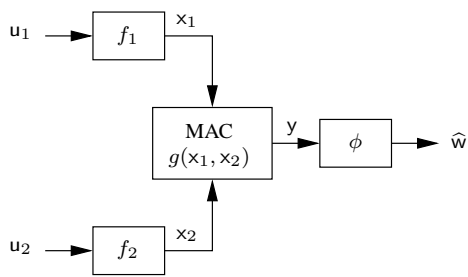}
    \end{center}
    \caption{Computation over a deterministic multiple-access channel.
    Each user $i$ has access to an independent message $\msf{u}_i$, and
    the receiver computes an estimate $\widehat{\msf{w}}$ of the target function
    $a(\msf{u}_1,\msf{u}_2)$ of those messages.}
    \label{Fig:Mac}
\end{figure}

A discrete, memoryless, deterministic two-user MAC consists of two \emph{input alphabets} $\mc{X}_1$ and $\mc{X}_2$,
an \emph{output alphabet} $\mc{Y}$, and a deterministic \emph{channel
function} $g\colon \mc{X}_1 \times \mc{X}_2 \to \mc{Y}$. Given channel inputs
$x_1, x_2$, the output of the MAC is 
\begin{equation*}
    y \defeq g(x_1,x_2).
\end{equation*}
Each transmitter $i\in\{1,2\}$ has access to an independent and
uniformly distributed \emph{message} $\msf{u}_i \in \mc{U}_i$. The
objective of the receiver is to compute a \emph{target function}
$a\colon \mc{U}_1 \times \mc{U}_2 \to \mc{W}$ of the user messages, see
Fig.~\ref{Fig:Mac}. 

Formally, each transmitter $i$ consists of an \emph{encoder}
$f_i\colon \mc{U}_i\to\mc{X}_i$ mapping the message $\msf{u}_i$ into the
channel input
\begin{equation*}
    \msf{x}_i \defeq f_i(\msf{u}_i).
\end{equation*}
The receiver consists of a \emph{decoder} $\phi\colon \mc{Y}\to\mc{U}$ mapping the
channel output $\msf{y}$ into an \emph{estimate} 
\begin{equation*}
    \hat{\msf{w}} \defeq \phi(\msf{y})
\end{equation*}
of the target function $a(\msf{u}_1,\msf{u}_2)$. The \emph{probability
of error} is 
\begin{equation*}
    \Pp\bigl(a(\msf{u}_1,\msf{u}_2)\neq \phi(\msf{y})\bigr).
\end{equation*}

\begin{remark}
    We point out that this differs from the ordinary communication setting,
    in which the decoder aims to recover both messages
    $(\msf{u}_1,\msf{u}_2)$. Instead, in the setting here, the decoder is
    not interested in $(\msf{u}_1,\msf{u}_2)$, but only in the value
    $a(\msf{u}_1,\msf{u}_2)$ of the target function.
\end{remark}

In the following, it will often be convenient to represent the target
function $a(\cdot,\cdot)$ and the channel $g(\cdot,\cdot)$ by their
corresponding matrices
$\bm{A}=(a_{u_1,u_2})\in\mc{W}^{\mc{U}_1\times\mc{U}_2}$ and
$\bm{G}=(g_{x_1,x_2})\in\mc{Y}^{\mc{X}_1\times\mc{X}_2}$, respectively.
In other words,
\begin{align*}
    a_{u_1,u_2} & = a(u_1,u_2)\in\mc{W}, \\
    g_{x_1,x_2} & = g(x_1,x_2)\in\mc{Y}.
\end{align*}
For $n\in\N$, denote by $\bm{G}^{\otimes n}$ the $n$-fold use of the
same channel matrix $\bm{G}$. In other words, the matrix
$\bm{G}^{\otimes n}$ describes the actions of the (memoryless) channel
$\bm{G}$ on the sequence
\begin{equation*}
    \bigl(
    (x_1[1],x_2[1]), 
    (x_1[2],x_2[2]), 
    \ldots,
    (x_1[n],x_2[n])
    \bigr)
\end{equation*}
of length $n$ of channel inputs.

\begin{definition}
    A pair $(\bm{A}, \bm{G})$ of target and channel functions is
    \emph{$\delta$-feasible}, if there exist encoders $f_1, f_2$ and a
    decoder $\phi$ computing the target function $\bm{A}$ over $\bm{G}$
    with probability of error at most $\delta$.
\end{definition}

\begin{remark}
    We will often consider pairs $(\bm{A},\bm{G}^{\otimes n})$, in which
    case the definition of $\delta$-feasibility allows for coding over
    $n$ uses of the channel $\bm{G}$. 
\end{remark}

Without loss of generality, we assume that the target function $\bm{A}$
has no two identical rows or two identical columns, since we could
otherwise simply eliminate one of them. For ease of exposition, we will
focus on the case
\begin{align*}
    \mc{U}_1 & = \mc{U}_2 = \mc{U},\\
    \mc{X}_1 & = \mc{X}_2 = \mc{X}.
\end{align*}
To simplify notation, we assume without loss of generality that
\begin{align*}
    \mc{U} & = \{0,1,\ldots, U-1\}, \ \ \
    \mc{X}  = \{0,1,\ldots, X-1\},\\
    \mc{W} & = \{0,1,\ldots, W-1\}, \ \ \hspace{.02in}
    \mc{Y}  = \{0,1,\ldots, Y-1\}.
\end{align*}
Finally, to avoid trivial cases, we assume that all cardinalities are
strictly bigger than one, and that $W\leq U^2$.

We denote by $\mc{A}(U,W)$ the collection of all target functions
$a\colon \mc{U}\times\mc{U}\to\mc{W}$. Similarly, we denote by $\mc{G}(X,Y)$
the collection of all channels $g\colon \mc{X}\times\mc{X}\to\mc{Y}$.  The
next example introduces several target functions $\bm{A}$ and channels
$\bm{G}$ that will be used to illustrate results in the remainder of the
paper. 

\begin{example}
    \label{eg:FnMAC}
    We start by introducing four target functions $a(\cdot,\cdot)$.
    \begin{itemize}
        \item Let $\mc{W} = \mc{U} \times \mc{U}$. The
            \emph{identity} target function is
            \begin{equation*}
                a(u_1, u_2) \defeq (u_1, u_2)
            \end{equation*} 
            for all $u_1, u_2 \in \mc{U}$. Since we will refer to the
            identity target function repeatedly, we will denote it by
            the symbol $\bm{A}_I$. 
        \item Let $\mc{W} = \{0, 1\}$. The \emph{equality} target function is
            \begin{equation*}
                a(u_1, u_2) \defeq
                \begin{cases}
                    1, & \text{if $u_1 = u_2$} \\
                    0, & \text{otherwise}
                \end{cases}
            \end{equation*}
            for all $u_1, u_2 \in  \mc{U}$.
        \item Let $\mc{W} = \{0,1\}$. The \emph{greater-than} target
            function is 
            \begin{equation*}
                a(u_1, u_2) \defeq
                \begin{cases}
                    1, & \text{if $u_1 > u_2$} \\
                    0, & \text{otherwise}
                \end{cases}
            \end{equation*}
            for all $u_1, u_2 \in  \mc{U}$.
        \item A \emph{random} target function corresponds to the matrix
            $\bm{\msf{A}}$ being a random variable, with each entry
            chosen independently and uniformly over $\mc{W}$. The matrix
            $\bm{\msf{A}}$ is generated before communication begins and
            is known at both the transmitters and at the receiver.
    \end{itemize}

    We now introduce three channels $g(\cdot,\cdot)$.
    \begin{itemize}
        \item Let  $\mc{X} = \{0,1\}$ and
            $\mc{Y} = \{0, 1, 2\}$. The \emph{binary
            adder} MAC is given by 
            \begin{equation*}
                g(x_1, x_2) \defeq x_1 + x_2
            \end{equation*}
            for all $x_1, x_2 \in \mc{X}$, and where $+$ denotes 
            ordinary addition.
        \item Let  $\mc{X} = \{0,1\}$ and $\mc{Y} = \{0,
            1\}$. The \emph{Boolean $\lor$} or \emph{Boolean OR} MAC is 
            \begin{equation*} 
                g(x_1, x_2) \defeq
                \begin{cases} 
                    0, & \text{if $x_1 = x_2 = 0$} \\
                    1, & \text{otherwise}
                \end{cases} 
            \end{equation*}
            for all $x_1, x_2 \in \mc{X}$.  
        \item A \emph{random}
            channel corresponds to the matrix $\bm{\msf{G}}$ being a
            random variable, with each entry chosen independently and
            uniformly over $\mc{Y}$. The matrix $\bm{\msf{G}}$ is
            generated before communication begins and is known at 
            both the transmitters and at the receiver.
    \end{itemize}
\end{example}

The emphasis in this paper is on the asymptotic behavior for large
function domains, i.e., as $U\to\infty$. We allow the other cardinalities
$X(U)$, $Y(U)$ and $W(U)$ to scale as a function of $U$. We use the
notation
\begin{equation*}
    X(U) \dotleq U^{a}
\end{equation*}
for the relation
\begin{equation*}
    \limsup_{U\to\infty}\frac{\log(X(U))}{\log(U)} \leq a
\end{equation*}
and analogously for $\dotl$. Similarly, we use
\begin{equation*}
    X(U) \dotgeq U^{a}
\end{equation*}
for the relation
\begin{equation*}
    \liminf_{U\to\infty}\frac{\log(X(U))}{\log(U)} \geq a
\end{equation*}
and analogously for $\dotg$. Finally,
\begin{equation*}
    X(U) \doteq U^{a}
\end{equation*}
is short hand for
\begin{equation*}
    X(U) \dotleq U^{a} \quad\text{ and }\quad X(U) \dotgeq U^{a}.
\end{equation*}
For example, $X(U) \doteq U^a$ is equivalent\footnote{Note that the notation $f(U)$ is $o(1)$ as $U\to\infty$ stands for $\lim_{U\to\infty}f(U)=0$.} to $X(U) = U^{a\pm
o(1)}$ as $U\to\infty$. With slight abuse of notation, we will write
$X(U) \dotl U^\infty$ to mean that $X(U) \dotleq U^\eta$ for \emph{some}
finite $\eta$. 

Throughout this paper, we are interested in efficient computation of the
target function $a(\cdot,\cdot)$ over the channel $g(\cdot,\cdot)$. In
Theorems~\ref{Thm:IdentityOneUse} and \ref{Thm:balanced} only a single
use of the channel is permitted, and efficiency is expressed in terms of
the required cardinalities $X(U)$ and $Y(U)$ of the channel alphabets as
a function of $U$. In Theorems \ref{Thm:IdentityMultipleUse} and
\ref{Thm:converse_n}, multiple uses of the channel are allowed, and
efficiency is then naturally expressed in terms of the number of
required channel uses $n(U)$ as a function of $U$.

Finally, all results are stated in terms of the fraction of channels (in
Theorems \ref{Thm:IdentityOneUse} and \ref{Thm:balanced}) or target
functions (in Theorem \ref{Thm:converse_n}) for which successful
computation is possible. The proofs of all the theorems are based on
probabilistic methods by using a uniform distribution over choices of
channel $g(\cdot,\cdot)$ or target functions $a(\cdot,\cdot)$.

\section{Main Results}
\label{sec:main}

\begin{figure}[ht]
    \begin{center}
        \hspace{-0.5cm}\input{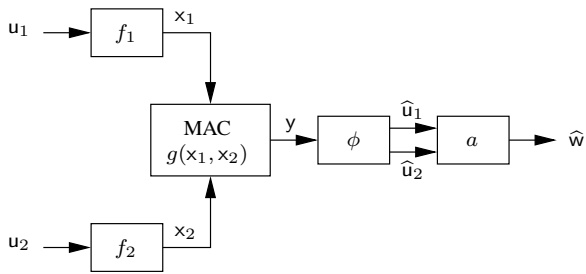}
    \end{center}
    \caption{Separation-based scheme computing the function
    $a(\cdot,\cdot)$ over the MAC $g(\cdot,\cdot)$. The receiver first
    decodes the original messages $(\widehat{\msf{u}}_1,\widehat{\msf{u}}_2)$
    and then evaluates the desired target function
    $a(\widehat{\msf{u}}_1,\widehat{\msf{u}}_2)$.}
    \label{Fig:Separation}
\end{figure}

Let $\bm{A}_I\in\mc{A}(U,U^2)$ be the identity target function
introduced in Example~\ref{eg:FnMAC}, and let $\bm{G}$ be an arbitrary
channel matrix. Consider any other target function
$\bm{A}\in\mc{A}(U,W)$ over the same domain $\mc{U}\times\mc{U}$, but
with possibly different range $\mc{W}$. Assume $(\bm{A}_I,\bm{G})$ is
$\delta$-feasible. Then $(\bm{A},\bm{G})$ is also $\delta$-feasible,
since we can first compute $\bm{A}_I$ (and hence $\widehat{\msf{u}}_1$
and $\widehat{\msf{u}}_2$) over the channel $\bm{G}$ and then simply
apply the function $\bm{A}$ to the recovered messages
$\widehat{\msf{u}}_1$ and $\widehat{\msf{u}}_2$. This architecture,
separating the computation task from the communication task, is
illustrated in Fig.~\ref{Fig:Separation}.

As a concrete example, let $\bm{A}$ be the greater-than target
function introduced in Example~\ref{eg:FnMAC}. The range
$\mc{W}=\{0,1\}$ of $\bm{A}$ has cardinality two. On the
other hand, the identity function $\bm{A}_I$ has range
$\mc{U}\times\mc{U}$ of cardinality $U^2$. In other words, for
large $U$, the identity target function is considerably more complicated
than the greater-than target function. As a result, one might expect
that the separation-based architecture in Fig.~\ref{Fig:Separation} is
highly suboptimal in terms of the computation efficiency as described in
Section \ref{sec:problem}. As the main result of this paper, we prove
that this intuition is wrong in most cases.  Instead, we show that for
most pairs $(\bm{A},\bm{G})$ of target function and MAC,
separation of computation and communication is close to optimal.

We discuss the single channel-use case in Section~\ref{sec:main_single},
and the $n$ channel-uses case in Section~\ref{sec:main_multiple}. 

\subsection{Single Channel Use ($n = 1$)}
\label{sec:main_single}

In this section, we will focus on the case where the target function
needs to be computed using just one use of the channel. The natural
value of the upper bound on the probability of error is $\delta=0$ in
this case. In other words, we will be interested in $0$-feasibility.

We start by deriving conditions under which computation of the identity
target function over a MAC
is feasible. Equivalently, these conditions guarantee that \emph{any}
target function with same domain cardinality $U$ can be computed over a
MAC by separating communication and computation as discussed above.

\begin{theorem}
    \label{Thm:IdentityOneUse}
    Let $\bm{A}_I\in\mc{A}(U,U^2)$ be the identity target function, and assume
    \begin{subequations}
        \label{eq:identity1}
        \begin{align}
            \label{eq:identity1a}
            X(U) &\dotg U , \\
            \label{eq:identity1b}
            Y(U) &\dotg U^3 .
        \end{align}
    \end{subequations}
    Then,
    \begin{equation*}
        \lim_{U\to\infty} \frac{\big\lvert\bigl\{\bm{G}\in\mc{G}(X(U),Y(U)):
        (\bm{A}_I, \bm{G}) \text{ is $0$-feasible} \bigr\}\big\rvert}
        {\lvert\mc{G}(X(U),Y(U))\rvert} = 1. 
    \end{equation*}
\end{theorem}

The proof of Theorem~\ref{Thm:IdentityOneUse} is reported in
Section~\ref{sec:proofs_identity}. Recall that $\mc{G}(X,Y)$ is the
collection of all channels $\bm{G}$ of dimension $X\times X$ and range
of cardinality $Y$. Theorem~\ref{Thm:IdentityOneUse} (together with the
separation approach discussed earlier) thus roughly implies that any
target function with a domain of cardinality $U$ can be computed over most
MACs of input cardinality $X(U)$ of order at least $U$ and output
cardinality $Y(U)$ of order at least $U^3$.  The precise meaning of
``most'' is that the fraction of channels $\bm{G}$ in $\mc{G}(X,Y)$ for which the
statement holds goes to one as $U\to\infty$. A look at the proof of the
theorem shows that the convergence to this limit is, in fact,
exponentially fast. In other words, the fraction of channels for which
the theorem fails to hold is exponentially small in the domain cardinality $U$. 
  
Since the achievable scheme is separation based, this conclusion holds
regardless of the cardinality $W(U)$ of the range of the target
function. Similarly, since it is clear that the channel input has to
have at least cardinality $X(U)$ of order $U$ for successful
computation, we see that the condition on $X(U)$ in
Theorem~\ref{Thm:IdentityOneUse} is not a significant restriction.  What
is significant, however, is the restriction that $Y(U)$ is at least of
order $U^3$. The next result shows that this restriction on $Y(U)$ is
essentially also necessary.

Before we state the theorem, we need to introduce one more concept.
\begin{definition}
    Consider a target function $a\colon \mc{U} \times \mc{U} \to \mc{W}$. For a
    set $\widetilde{\mc{W}}\subset \mc{W}$, consider
    \begin{equation*}
        a^{-1}(\widetilde{\mc{W}}) 
        \defeq \{ (u_1, u_2)\in\mc{U}\times\mc{U}: a(u_1, u_2) \in \widetilde{\mc{W}} \}.
    \end{equation*}  
    For $c \in (0, 1/2]$, the target function $a(\cdot,\cdot)$ is said to be
    \emph{$c$-balanced} if there exist a partition $\mc{W}_1, \mc{W}_2$
    of $\mc{W}$ such that 
    \begin{equation*}
        \card{ a^{-1}(\mc{W}_i) } \geq c\cdot U^2 
    \end{equation*}  
    for all $i\in\{1,2\}$.
\end{definition}

Most functions are $c$-balanced for any $c< 1/3$ and $W(U)$ as long as
$U$ is large enough.  Indeed, choosing $\mc{W}_1 = \{0,\ldots,
\floor{W(U)/2}-1\}$ and $\mc{W}_2 = \{\floor{W(U)/2},\ldots, W(U)-1\}$
shows that
\begin{equation}
    \label{eq:balancedfraction}
    \lim_{U\to\infty} 
    \frac{\big\lvert\bigl\{\bm{A}\in\mc{A}(U,W(U)):
    \bm{A} \text{ is $1/3$-balanced} \bigr\}\big\rvert}
    {\lvert\mc{A}(U,W(U))\rvert}
    = 1,
\end{equation}
where we recall that $\mc{A}(U,W)$ denotes the collection of all target
functions $\bm{A}$ of dimension $U\times U$ and range of cardinality
$W$. In fact, the convergence in \eqref{eq:balancedfraction} is again
exponentially fast\footnote{This follows directly from results on the
convergence of empirical distributions.} in $U$.  Moreover, many
functions of specific interest are balanced.

\begin{example}
    \label{eg:balanced}
    Consider the target functions introduced in Example~\ref{eg:FnMAC}.
    \begin{itemize}
        \item 
            The identity and the greater-than target functions are
            $c$-balanced for any constant $c<1/2$ and $U$ large enough.
        \item
            The equality target function is \emph{not} $c$-balanced for
            any constant $c>0$ as $U\to\infty$. Indeed, since $W(U)=2$
            in this case, the only choice (up to labeling) is to set
            $\mc{W}_1 = \{0\}$ and $\mc{W}_2 = \{1\}$. Then
            $\card{a^{-1}(\mc{W}_1)} = U^2-U$ and
            $\card{a^{-1}(\mc{W}_2)} = U$, which is not $c$-balanced for
            any constant $c>0$ as $U\to\infty$.
    \end{itemize}
\end{example}

We have the following converse result to
Theorem~\ref{Thm:IdentityOneUse} for balanced target functions. 
\begin{theorem}
    \label{Thm:balanced}
    Fix a constant $c\in(0,1/2]$ independent of $U$. Assume $W(U)\geq 2$
    and
    \begin{subequations}
        \label{eq:balanced0}
        \begin{align}
            \label{eq:balanced0a}
            X(U) & \dotl U^\infty, \\
            \label{eq:balanced0b}
            Y(U) & \dotl U^3.
        \end{align}
    \end{subequations}
    Let $\bm{A}\in\mc{A}(U,W(U))$ be any $c$-balanced target function.
    Then 
    \begin{equation*}
        \lim_{U\to\infty} \frac{\big\lvert\bigl\{\bm{G}\in\mc{G}(X(U),Y(U)):
        (\bm{A}, \bm{G}) \text{ is $0$-feasible} \bigr\}\big\rvert}
        {\lvert\mc{G}(X(U),Y(U))\rvert} = 0. 
    \end{equation*}
\end{theorem}

The proof of Theorem~\ref{Thm:balanced} is reported in
Section~\ref{sec:proofs_balanced}.  Recall that the notation $X(U)\dotl
U^\infty$ is used to indicate that $X(U)$ grows at most polynomially in
$U$---an assumption that is quite mild. Thus, Theorem~\ref{Thm:balanced}
roughly states that regardless of the value of $W(U)$, if the
cardinality $Y(U)$ of the channel output is order-wise less than $U^3$,
then any balanced target function with a range of cardinality $W(U)$
cannot be computed over most MACs. Here the precise meaning of ``most''
is again that the fraction of channel matrices with at most $Y(U)$
channel outputs for which successful computation is possible converges
to zero, and a look at the proof reveals again that this convergence is,
in fact, exponentially fast in $U$. 

Comparing this to Theorem~\ref{Thm:IdentityOneUse}, we see that the same
scaling of $Y(U)$ allows computation of a target function using a
separation based scheme (i.e., by first recovering the two messages
$(\hat{\msf{u}}_1,\hat{\msf{u}}_2)$ and then applying the target
function to compute the estimate $\hat{\msf{w}} =
a(\hat{\msf{u}}_1,\hat{\msf{u}}_2)$). Thus, for the computation of a
given balanced function over most MACs, separation of computation and
communication is essentially optimal. Moreover, since most functions are
balanced by \eqref{eq:balancedfraction}, the same also holds for most
pairs $(\bm{A},\bm{G})$ of target and channel functions.
  
\begin{example}
    \label{eg:greater} 
    Let $\bm{A}$ be the greater-than target function of domain $U\times
    U$ introduced in Example~\ref{eg:FnMAC}. Note that this target
    function has range of cardinality $W(U)=2$, i.e., $\bm{A}$ is
    binary. From Example~\ref{eg:balanced}, we know that $\bm{A}$ is
    balanced for any constant $c< 1/2$ and $U$ large enough. Thus
    Theorem~\ref{Thm:balanced} applies, showing that, for large $U$ and
    most MACs $\bm{G}$, separation of computation and
    communication is essentially optimal.

    Observe that the receiver is interested in only a \emph{single} bit
    of information about $(\msf{u}_1,\msf{u}_2)$. Nevertheless, the
    structure of the greater-than target function is complicated enough
    that, in order to recover this single bit, the decoder is
    essentially forced to learn $(\msf{u}_1,\msf{u}_2)$ itself. In other
    words, in order to compute the single desired bit, communication of
    $2\log(U)$ message bits is essentially necessary. 
\end{example}

Theorem~\ref{Thm:balanced} is restricted to balanced functions. Even
though only a vanishingly small fraction of target functions is not
balanced, it is important to understand this restriction. We illustrate
this through the following example.

\begin{example}
    \label{eg:equality}
    Assume $W(U)=2$ and
    \begin{subequations}
        \label{eq:equality0}
        \begin{align}
            X(U) & \dotg U, \\
            Y(U) & \dotg U.
        \end{align}
    \end{subequations}
    Let $\bm{A}_{=}\in\mc{A}(U,2)$ be the equality target function
    introduced in Example~\ref{eg:FnMAC}. Then 
    \begin{equation}
        \label{eq:equality}
        \lim_{U\to\infty} \frac{\big\lvert\bigl\{\bm{G}\in\mc{G}(X(U),Y(U)):
        (\bm{A}_=, \bm{G}) \text{ is $0$-feasible} \bigr\}\big\rvert}
        {\lvert\mc{G}(X(U),Y(U))\rvert} = 1. 
    \end{equation}

    The proof of the above statement is reported in
    Section~\ref{sec:proofs_equality}.  This result shows that the
    equality function can be computed over a large fraction of MACs with
    output cardinality $Y(U)$ of order at least $U$.  This contrasts
    with output cardinality $Y(U)$ of order $U^3$ that is required for
    successful computation of balanced functions in
    Theorem~\ref{Thm:balanced}. Recall from Example~\ref{eg:balanced}
    that the equality target function is \emph{not} $c$-balanced for any
    $c>0$ and $U$ large enough. Thus, \eqref{eq:equality} does not
    contradict Theorem~\ref{Thm:balanced}. It does, however, show that
    for unbalanced functions separation of communication and computation
    can be suboptimal.
\end{example}

\subsection{Multiple Channel Uses ($n\geq 1$)} 
\label{sec:main_multiple}

In this section, we allow multiple uses of the MAC. Our emphasis will
again be on the asymptotic behavior for large function domains
$U\to\infty$. However, in this section we keep the MAC $g(\cdot,
\cdot)$, and hence also the cardinalities of the channel domain $\mc{X}$
and channel range $\mc{Y}$, fixed. Instead, we characterize the minimum
number $n=n(U)$ of channel uses required to compute the target function. 

We begin by stating a result for the identity target function introduced
in Example~\ref{eg:FnMAC}. Equivalently, this result applies to
\emph{any} target function (with same domain cardinality $U$) by using a
scheme separating communication and computation. Let $H(\msf{x})$ denote
the entropy of a random variable $\msf{x}$.

\begin{theorem}
    \label{Thm:IdentityMultipleUse} 
    Fix a constant $\delta > 0$ independent of $U$, and assume that $X$
    and $Y$ are constant. Let $\bm{A}_I\in\mc{A}(U,U^2)$ be the
    identity target function, and let $\bm{G}\in\mc{G}(X,Y)$ be any
    MAC.  Consider any joint distribution of the form $p(q)p(x_1 |
    q)p(x_2 | q) p(y | x_1, x_2)$, where $p(y|x_1,x_2)$  is specified
    by the channel function $\bm{G}$. Then, for any $n(U)$ satisfying
    \begin{subequations}
        \label{Eq:IdentityConstraints}
        \begin{align}
            U^2 &\dotl 2^{n(U)H\left(\msf{y} | \msf{q}\right)}, \\
            U &\dotl 2^{n(U)H(\msf{y} | \msf{x}_1, \msf{q})}, \\
            U &\dotl 2^{n(U)H(\msf{y} | \msf{x}_2, \msf{q})},
        \end{align}
    \end{subequations}
    $(\bm{A}_I, \bm{G}^{\otimes n(U)})$ is $\delta$-feasible
    for $U$ large enough.
\end{theorem}

The result follows directly from the characterization of the achievable
rate region for ordinary communication over the MAC, see for example
\cite[Theorem~14.3.3]{cover91}.  Using separation,
Theorem~\ref{Thm:IdentityMultipleUse} implies that, for large enough
$U$, any target function of domain cardinality $U$ can be reliably
computed over $n(U)$ uses of an MAC $\bm{G}$ as long as it satisfies the
constraints in \eqref{Eq:IdentityConstraints}. The next result states
that for most functions these restrictions on $n(U)$ are essentially
also necessary.

\begin{theorem}
    \label{Thm:converse_n}
    Assume that\footnote{Note that the notation $W(U) \geq \omega(1)$
    as $U\to\infty$ stands for $\lim_{U\to\infty}W(U)=\infty$.}
    \begin{equation*}
        W(U) \geq \omega(1)
    \end{equation*}
    as $U\to\infty$, that $0 < \delta \le 1 / (2\ln(W(U)))$, and that $X$ and $Y$ are constant.
    Let $\bm{G}\in\mc{G}(X,Y)$ be any MAC. Then, for any
    $n(U)$ satisfying
    \begin{equation*}
        \lim_{U\to\infty} \frac{\big\lvert\bigl\{\bm{A}\in\mc{A}(U,W(U)):
        (\bm{A}, \bm{G}^{\otimes n(U)}) \text{ is $\delta$-feasible} \bigr\}\big\rvert}
        {\lvert\mc{A}(U,W(U))\rvert} > 0 , 
    \end{equation*}
    we must have 
    \begin{align*}
        U^2 &\dotleq 2^{n(U)H\left(\msf{y} | \msf{q}\right)}, \\
        U &\dotleq 2^{n(U)H(\msf{y} | \msf{x}_1, \msf{q})}, \\
        U &\dotleq 2^{n(U)H(\msf{y} | \msf{x}_2, \msf{q})}
    \end{align*}
    for some joint distribution of the form $p(q)p(x_1 | q)p(x_2 | q)
    p(y | x_1, x_2)$, where $p(y|x_1,x_2)$ is specified by the channel
    function $\bm{G}$. 
\end{theorem}

The proof of Theorem~\ref{Thm:converse_n} is presented in
Section~\ref{sec:proofs_n}. Recall that $\mc{A}(U,W)$ denotes the
collection of all target functions $\bm{A}$ of dimension $U\times U$ and
range of cardinality $W$. Together,
Theorems~\ref{Thm:IdentityMultipleUse} and \ref{Thm:converse_n} thus
show that, for any deterministic MAC and most target functions, the
smallest number of channel uses $n^\star(U)$ that enables reliable
computation is of the same order as that needed for the identity
function. Moreover, they show that for most such pairs, separation of
communication and computation is essentially optimal even if we allow
multiple uses of the channel and nonzero error probability. Here the
precise meaning of ``most'' is that the statement holds for all but a
vanishing fraction of functions. Moreover, the proof of the theorem
shows again that this fraction is, in fact, exponentially small in $U$. 

\begin{example}
    \label{eg:n}
    Let $\bm{G}$ be the binary adder MAC introduced in
    Example~\ref{eg:FnMAC}. Define
    \begin{equation*}
        H^\star(\bm{G})  \defeq \max_{\msf{x}_1,\msf{x}_2}  H\bigl(g(\msf{x}_1,\msf{x}_2)\bigr) = 3/2 , 
    \end{equation*}
    where the maximization is over all independent random variables
    $\msf{x}_1,\msf{x}_2$ taking values in the channel input alphabet
    $\mc{X}$. $H^\star(\bm{G})$ denotes the maximum entropy that can be
    induced at the channel output. For the binary adder MAC $\bm{G}$, it
    follows from \eqref{Eq:IdentityConstraints} in
    Theorem~\ref{Thm:IdentityMultipleUse} that the identity function can
    be reliably computed over $\bm{G}$ if the number of channel uses
    $n(U) \ge 2\log U / H^\star(\bm{G}) = 4\log U / 3$.  On the other
    hand, Theorem~\ref{Thm:converse_n} shows that for most functions
    $\bm{A}$ of domain $U\times U$ and range cardinality $W(U) =
    \log(U)$, the smallest number of channel uses $n^\star(U)$ required
    for reliable computation is of order $4\log(U)/3$.  Thus,
    near-optimal performance can be achieved by separating computation
    and communication. In other words, even though the receiver is only
    interested in $\log\log(U)$ function bits, it is essentially forced
    to learn the $2\log(U)$ message bits as well.

    This example also illustrates that the usual way of proving converse
    results based on the cut-set bound is not tight for most
    $(\bm{A},\bm{G})$ pairs. For example, \cite[Lemma~13]{nazer07} shows
    that for reliable computation we need to have
    \begin{equation*}
        n(U)H^\star(\bm{G}) \geq H(a(\msf{u}_1,\msf{u}_2))
    \end{equation*}
    where $H(\cdot)$ denotes entropy. Since $\bm{A}$ has range of
    cardinality $W(U)$, we have
    \begin{equation*}
        H(a(\msf{u}_1,\msf{u}_2)) \leq \log(W(U)).
    \end{equation*}
    For $W(U)=\log(U)$ and $H^\star(\bm{G})=3/2$ as considered here, the
    tightest bound that can in the \emph{best case} be derived via the
    cut-set approach is thus
    \begin{equation*}
        n^\star(U) \geq \log(W(U))/H^\star(\bm{G}) = 2\log\log(U)/3.
    \end{equation*}
    However, we know that the correct scaling for $n^\star(U)$ is
    $4\log(U)/3$. Hence, the cut-set bound is loose by an unbounded
    factor as $U\to\infty$.
\end{example}

\section{Proofs}
\label{sec:proofs}

We now prove the main results. The proofs of
Theorems~\ref{Thm:IdentityOneUse} and \ref{Thm:balanced} are reported in
Sections~\ref{sec:proofs_identity} and \ref{sec:proofs_balanced}
respectively. The proof of \eqref{eq:equality} in
Example~\ref{eg:equality} is presented in
Section~\ref{sec:proofs_equality}. Finally, the proof of
Theorem~\ref{Thm:IdentityMultipleUse} is covered in
Section~\ref{sec:proofs_n}. We start by presenting some preliminary
observations in Section~\ref{sec:proofs_preliminaries}.

\subsection{Preliminaries}
\label{sec:proofs_preliminaries}

Recall our assumption that no two rows or two columns of the target
function $\bm{A}$ are identical. As a result, $\bm{A}$ can be computed
over the channel $\bm{G}$ with zero error, i.e., $(\bm{A},\bm{G})$ is
$0$-feasible, if and only if there exists a $U \times U$ submatrix (with
ordered rows and columns) $\bm{S}$ of $\bm{G}$ such that any two entries
$(u_1, u_2)$ and $(\tilde{u}_1, \tilde{u}_2)$ with $a_{u_1, u_2} \neq
a_{\tilde{u}_1,\tilde{u}_2}$ satisfy $s_{u_1,u_2} \neq
s_{\tilde{u}_1,\tilde{u}_2}$, see Fig.~\ref{Fig:Scheme}. 

\begin{figure}[ht]
    \begin{center}
        \vspace{9pt}
        \scalebox{.444}{\hspace{-0.2cm}\input{figs/Scheme.pstex_t}}
    \end{center}
    \caption{Structure of a zero-error computation scheme over a MAC.
    The target function $\bm{A}$ corresponds to the equality function,
    the MAC matrix $\bm{G}$ corresponds to the Boolean $\lor$ MAC, and
    $\bm{G}^{\otimes 2}$ denotes the $2$-fold  use of channel
    $\bm{G}$. While the function $\bm{A}$ cannot be computed over
    $\bm{G}$ in one channel use, it can be computed in two channel uses
    by assigning the channel input $01$ to user message $0$ and $10$ to
    user message $1$. The corresponding ordered submatrix $\bm{S}$ of
    $\bm{G}$ is indicated in bold lines in the figure.}
    \label{Fig:Scheme}
\end{figure}

On the other hand, this is not necessary if a probability of error
$\delta > 0$ can be tolerated. As an example, consider the equality
function. For any positive $\delta$, a trivial decoder that always
outputs $0$ computes the equality function with probability of error
$1/U$. As $U\to\infty$, the probability of error is eventually less than
$\delta$. This motivates the following definition.

Given a target function $a\colon \mc{U} \times \mc{U} \to
\mc{W}$, a function   $a_\delta\colon \mc{V}_1 \times \mc{V}_2 \to
\mc{W}$ with 
\begin{align*}
    V_1 & \defeq \card{\mc{V}_1} \leq U, \\
    V_2 & \defeq \card{\mc{V}_2} \leq U
\end{align*}
is said to be a \emph{$\delta$-approximation} of $a(\cdot,\cdot)$ if
there exist two mappings $f_1\colon \mc{U}\to\mc{V}_1$ and
$f_2\colon \mc{U}\to\mc{V}_2$ such that 
\begin{align}
    \label{Eqn:ApproxFunction}
    \big\lvert\bigl\{(u_1, u_2) \in \mc{U}\times\mc{U} : 
    a(u_1, u_2) \neq a_{\delta}\bigl(f_1(u_1), f_2(u_2)\bigr)\bigr\}
    \big\rvert 
    \leq \delta U^2.   
\end{align}
In words, the target function $a(\cdot,\cdot)$ is equal to the approximation function $a_{\delta}(\cdot,\cdot)$ for at least a 
$(1-\delta)$ fraction of all message pairs. As before, a
$\delta$-approximation function $a_{\delta}$ can be represented by a
$V_1 \times V_2$ matrix $\bm{A}_{\delta}$. We have the following
straightforward relation.

\begin{lemma}
    \label{Lemma:Translation}
    Consider a target function $\bm{A}$ and MAC $\bm{G}$. If
    $(\bm{A},\bm{G})$ is $\delta$-feasible, then there exists a
    $\delta$-approximation $\bm{A}_{\delta}$ of $\bm{A}$ such that
    $(\bm{A}_\delta,\bm{G})$ is $0$-feasible.
\end{lemma}
\begin{IEEEproof}
    Let $f_1,f_2$ and $\phi$ be the encoders and the decoder achieving
    probability of error at most $\delta$ for $(\bm{A},\bm{G})$. Let
    $\mc{V}_i$ be the range of $f_i$, and set
    \begin{equation*}
        a_\delta(f_1(u_1),f_2(u_2)) 
        \defeq \phi\bigl(g(f_1(u_1),f_2(u_2))\bigr)
    \end{equation*}
    for all $u_1,u_2\in\mc{U}$. Then $a_\delta(\cdot,\cdot)$ is a
    $\delta$-approximation of $a(\cdot,\cdot)$, and $(a_\delta,g)$ is
    $0$-feasible.
\end{IEEEproof}

We will make frequent use of the Chernoff bound, which we recall here
for future reference. Let $\msf{z}_1, \msf{z}_2,\ldots,\msf{z}_N$ be
independent random variables, and let 
\begin{equation*}
    \msf{z} \defeq \sum_{i=1}^{N} \msf{z}_i.
\end{equation*}
By Markov's inequality,
\begin{equation}
    \label{Eqn:Chernoff1}
    \Pp(\msf{z} > b) 
    < \min_{t > 0} \exp(-tb)\prod_{i=1}^N \E\bigl(\exp(t\msf{z}_i)\bigr)
\end{equation}
Assume furthermore that each $\msf{z}_i$ takes value in $\{0,1\}$, and
set
\begin{equation*}
    \mu \defeq \E(\msf{z}).
\end{equation*}
Then, for any $\gamma > 0$, 
\begin{align}
    \label{Eqn:Chernoff2}
    \Pp\bigl(\msf{z} > (1 + \gamma)\mu\bigr) 
    & < \Bigl( \frac{e^\gamma}{(1 + \gamma)^{(1 + \gamma)}}\Bigr)^{\mu}, \\
    \shortintertext{and, for $0 < \gamma \leq 1$,}\nonumber\\
    \label{Eqn:Chernoff3}
    \Pp\left(\msf{z} < (1 - \gamma)\mu\right) 
    & < \exp(-\mu\gamma^2/2),
\end{align}
see for example \cite[Theorem~4.1, Theorem~4.2]{motwani95}.

\subsection{Proof of Theorem~\ref{Thm:IdentityOneUse}}
\label{sec:proofs_identity}

A scheme can compute the identity target function with zero error if and
only if the channel output corresponding to any two distinct pairs of
user messages is different. In what follows, we will show that such a
scheme for computing the identity target function over \emph{any} MAC
$\bm{G}$ exists whenever the elements of $\bm{G}$ take at least
$X^2(U)-X(U)+1$ distinct values in $\mc{Y}$. We then argue that, if the
assumptions on $X(U)$ and $Y(U)$ in \eqref{eq:identity1} are satisfied,
a \emph{random} MAC $\bm{\msf{G}}$ (as introduced in
Example~\ref{eg:FnMAC}) of dimension $X(U) \times X(U)$ has at least
$X^2(U)-X(U)+1$ distinct entries with high probability as $U\to\infty$.
Together, this will prove the theorem.

Note that \eqref{eq:identity1} implies that $X \geq 4U$ and $Y \geq
64e^3U^3$ for $U$ large enough. We will prove the result under these two
weaker assumptions on $X(U)$ and $Y(U)$. Since we can always choose to
ignore part of the channel inputs, we may assume without loss of
generality that $X(U) = 4U$. In order to simplify notation, we suppress
the dependence of $Y=Y(U)$ and $X=X(U)$ on $U$ in the remainder of this
and all other proofs.

\begin{figure}[ht]
    \begin{center}
        \scalebox{.5}{\input{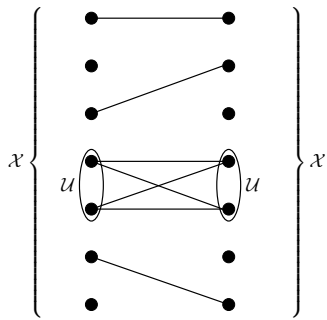}}
    \end{center}
    \caption{Bipartite graph $B$ representing the channel matrix
    $\bm{G}$. The left vertices correspond to the possible channel
    inputs at transmitter one, and the right vertices correspond to the
    possible channel inputs at transmitter two. Each edge of $B$
    corresponds to a distinct value in $\bm{G}$. Thus, the number
    of edges in $B$ is equal to the number of distinct channel
    outputs. The existence of a $U \times U$ complete subgraph $K_{U,U}$
    corresponds to the existence of two sets of channel inputs each of size
    $U$ such that all corresponding channel outputs are different. Thus,
    these sets of channel inputs can be used to compute the identity
    target function over $\bm{G}$ with zero error.}
    \label{Fig:Identity}
\end{figure}

Given an arbitrary MAC $\bm{G}$, create a bipartite graph $B$ as follows
(see Fig.~\ref{Fig:Identity}). Let the vertices of $B$ on each of the
two sides of the bipartite graph be $\mc{X}$. Now, consider a value
$y\in\mc{Y}$ appearing in $\bm{G}$. This $y$ corresponds to a collection
of vertex pairs $(x_1,x_2)$ such that $g_{x_1,x_2}=y$. Pick exactly one
arbitrary such vertex pair $(x_1,x_2)$ and add it as an edge to $B$. Repeat
this procedure for all values of $y$ appearing in $\bm{G}$. Thus, the
total number of edges in the graph $B$ is equal to the number of
distinct entries in the channel matrix $\bm{G}$. 

Observe that any complete $U \times U$ bipartite subgraph $K_{U,U}$ of
the bipartite graph $B$ corresponds to a computation scheme for
the identity function. Indeed, by construction each edge in $B$
corresponds to a different channel output. Hence by encoding $\mc{U}_1$
and $\mc{U}_2$ as the left and right vertices, respectively, of the
subgraph $K_{U,U}$, we can uniquely recover $(u_1,u_2)$ from the channel
output $g(u_1,u_2)$. 

This problem of finding a bipartite subgraph $K_{U,U}$ in the bipartite
graph $B$ is closely related to the Zarankiewicz problem, see for example \cite[Chapter VI]{bollobas04}.
Formally, the aim in the Zarankiewicz problem is to characterize
$Z_{b}(n)$, the smallest integer $m$ such that every bipartite graph
with $n$ vertices on each side and $m$ edges contains a subgraph
isomorphic to $K_{b,b}$. The K\H{o}v\'{a}ri-S\'{o}s-Tur\'{a}n theorem,
see for example \cite[Theorem~VI.2.2]{bollobas04}, states that 
\begin{equation}
    \label{eq:Zaran}
    Z_{b}(n) < (b-1)^{1/b} (n - b + 1)n^{1 - 1/b} + (b-1)n + 1.
\end{equation} 

Using \eqref{eq:Zaran}, we now argue that the bipartite graph $B$
defined above contains a complete $U \times U$ bipartite subgraph
$K_{U,U}$ if the number of edges in $B$ is at least $X^{2}-X+1$.  By
definition, $B$ contains $K_{U,U}$ if there are at least $Z_{U}(X)$
edges in $B$. By~\eqref{eq:Zaran},
\begin{align}
    \label{eq:Zaran2}
    Z_{U}(X) 
     &< (U - 1)^{1/U} (X - U + 1) X^{1 - 1/U} + (U -1)X+1  \nonumber\\
    & = X(X - 1) + 1 + X(X - U + 1)\biggl( \Bigl(\frac{U - 1}{X} \Bigr)^{1/U}\!\!\!\! - \frac{X - U}{X - U + 1} \biggr) 
    \biggr).
\end{align}
Using the inequality $(1-x)^n \ge 1 - nx$ for $x\in[0,1]$ and that
$X=4U$ by assumption, we have
\begin{align*}
    \Bigl( \frac{X - U}{X - U + 1} \Bigr)^{U} 
    & = \Bigl(1 - \frac{1}{X - U + 1}\Bigr)^U  \\
    & \geq 1 - \frac{U}{X - U + 1}  
     = \frac{2U + 1}{X-U+1} \\
    & \geq \frac{U - 1}{X}.
\end{align*}
Combining this with \eqref{eq:Zaran2} shows that
\begin{equation*}
    Z_U(X) <  X^2-X+1.
\end{equation*}

Thus we have shown that the identity target function can be computed
over \emph{any} channel $\bm{G}$ with $X=4U$ if it has at least
$X^2-X+1$ distinct entries. Consider now a \emph{random} channel
$\bm{\msf{G}}$. The next lemma shows that $\bm{\msf{G}}$ satisfies this condition with high probability as $X\to\infty$.
\begin{lemma}
    \label{Lemma:DistinctEntries}
    Let $\msf{N}$ be the number of distinct entries in the random channel
    matrix $\bm{\msf{G}}$, and assume $Y\geq e^3X^3$. Then
    \begin{equation*}
        \Pp(\msf{N} \geq X^2-X+1) \geq 1 - \exp(-(X-2))
    \end{equation*}
    for $X$ large enough.
\end{lemma}

The proof of  Lemma~\ref{Lemma:DistinctEntries} is
reported in Appendix~\ref{sec:appendix_distinct}.
Lemma~\ref{Lemma:DistinctEntries} shows that with probability at least
$1 - \exp(-(X-2))$ the identity target function can be computed with
zero error over the random MAC $\bm{\msf{G}}$. Since $X = 4U$ so that
$X\to\infty$ as $U\to\infty$, the statement of the theorem follows.
\hfill\IEEEQED

\subsection{Proof of Theorem~\ref{Thm:balanced}}
\label{sec:proofs_balanced}

Let $\mc{W}_1$ and $\mc{W}_2$ be the two sets in the definition of a
balanced function. For a MAC $\bm{G}$, a $U \times U$ ordered submatrix
$\bm{S}$ corresponds to a valid code for computing $\bm{A}$ with zero
error only if there are no common values between the entries in $\bm{S}$
corresponding to $a^{-1}(\mc{W}_1)$ and $a^{-1}(\mc{W}_2)$. Consider
then a random $\bm{\msf{G}}$ (as introduced in Example~\ref{eg:FnMAC})
and one such ordered submatrix $\bm{\msf{S}}$. Observe that the
selection of rows and columns of $\bm{\msf{G}}$ in $\bm{\msf{S}}$ is
fixed---the matrix $\bm{\msf{S}}$ is random because its entries are
derived from the random matrix $\bm{\msf{G}}$. Let $\msf{N}_1$ denote
the number of distinct values among the entries corresponding to
$a^{-1}(\mc{W}_1)$ in $\bm{\msf{S}}$. We have the following bound on
$\msf{N}_1$. 
    
\begin{lemma}
    \label{Lemma:DistinctEntries2}
    Assume $a^{-1}(\mc{W}_1)\geq cU^2$, and set
    \begin{equation*}
        N \defeq \min \{Y/3, cU^2 / 3\}.
    \end{equation*}
    Then
    \begin{equation*}
        \Pp(\msf{N}_1 < N) \leq \exp(-cU^{2}/6).
    \end{equation*}
\end{lemma} 

The proof of Lemma~\ref{Lemma:DistinctEntries2} is reported in
Appendix~\ref{sec:appendix_distinct2}.  The submatrix $\bm{\msf{S}}$
corresponds to a valid code for computing the target function $\bm{A}$
only if all the entries corresponding to $a^{-1}(\mc{W}_2)$ take values
from the $Y - \msf{N}_1$ channel outputs not present in the entries
corresponding to $a^{-1}(\mc{W}_1)$. Thus the probability of the
submatrix $\bm{\msf{S}}$ being a valid code for computing the target
function $\bm{A}$ is at most 
\begin{align}
    \label{eq:balanced1}
    \Pp(\bm{\msf{S}} \text{ is a  valid code for $\bm{A}$} )  &\leq \Pp(\msf{N}_1 < N) + \left( \frac{ Y - N}{Y} \right)^{\card{a^{-1}(\mc{W}_2)}} \nonumber\\
     &\stackrel{(a)}{\leq} \exp(- cU^{2}/6) + \exp(- N \card{a^{-1}(\mc{W}_2)}/Y) \nonumber\\
     &\stackrel{(b)}{\leq} \exp(- cU^{2}/6) + \exp(- N cU^2/Y) \nonumber\\
    &\stackrel{(c)}{\leq} 2\exp\bigl(- \min\{cU^{2}/6, c^2U^4/(3Y)\} \bigr),
\end{align}
where $(a)$ follows from Lemma~\ref{Lemma:DistinctEntries2} and $1 - x
\leq e^{-x}$, $(b)$ follows since $\bm{A}$ is $c$-balanced, and $(c)$
follows from the definition of $N$.

The pair $(\bm{A},\bm{\msf{G}})$ is $0$-feasible if and only if there
exists some valid ordered submatrix $\bm{\msf{S}}$ with dimension $U
\times U$ of the $X \times X$ channel matrix $\bm{\msf{G}}$. There are
at most $X^{2U}$ ways to choose the rows and columns of this submatrix.
Hence, from the union bound and~\eqref{eq:balanced1}, the probability
that $(\bm{A},\bm{\msf{G}})$ is $0$-feasible is at most 
\begin{align*}
    \Pp((\bm{A},\bm{\msf{G}}) \text{ is $0$-feasible} )
    & \leq X^{2U} \Pp(\bm{\msf{S}} \text{ is a valid code for $\bm{A}$} ) \nonumber\\
    & \leq X^{2U}\cdot 2\exp\bigl(- \min\{cU^{2}/6, c^2U^4/(3Y) \} \bigr) \nonumber\\
    & = \exp\bigl(- \bigl( \min\{cU^{2}/6, c^2U^4/(3Y)\}- 2U\ln(X) - \ln(2) \bigr) \bigr).
\end{align*}
Now, \eqref{eq:balanced0} implies that $X \leq U^{m}$ and $Y \leq c^2U^3
/ (12m\ln(U))$ for some finite positive $m$ and $U$ large enough. Hence,
\begin{equation*}
    \lim_{U\to\infty} \Pp((\bm{A},\bm{\msf{G}}) \text{ is $0$-feasible} ) = 0,
\end{equation*}
as needed to be shown. \hfill\IEEEQED

\subsection{Proof of \eqref{eq:equality} in Example~\ref{eg:equality}}
\label{sec:proofs_equality}

A scheme computes the equality target function $\bm{A}_{=}$ with zero error if and
only if the channel outputs corresponding to the set of message pairs
$\{(u,u): u \in \mc{U}\}$ are all distinct from those corresponding to
the message pairs $\{(u_1, u_2): u_1 \neq u_2\}$. In what follows, we
will first show that this is guaranteed if the channel matrix $\bm{G}$
satisfies certain properties. We then argue that a random channel matrix
$\bm{\msf{G}}$ (as introduced in Example~\ref{eg:FnMAC}) satisfies these
properties with high probability.

From \eqref{eq:equality0}, we can assume that $X \geq 200U\ln(U)$ and $Y
\geq 16U$ for $U$ large enough. We will prove the result under these two
weaker conditions. Since the encoders can always choose to ignore some
of the channel inputs, we can assume without loss of generality that $X
= 200U\ln(U)$. Throughout this proof, we denote by $k$ the largest
integer such that $Y/k \geq 16U$. Note that this implies
\begin{equation}
    \label{eq:k}
    16U \leq Y/k < 32U.
\end{equation}

\begin{figure}[ht]
    \begin{center}
        \scalebox{.5}{\input{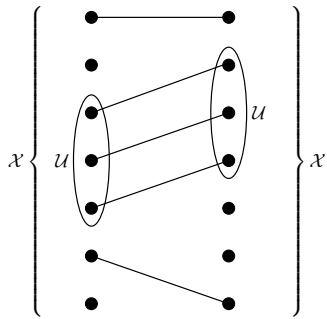}}
    \end{center}
    \caption{Bipartite graph $B$ representing the channel matrix
    $\bm{G}$. The left vertices correspond to the possible channel
    inputs at transmitter one, and the right vertices correspond to the
    possible channel inputs at transmitter two. For some fixed subset
    $\tilde{\mc{Y}}\subset\mc{Y}$ of cardinality $k$, the graph contains
    an edge between two vertices $x_1,x_2$ if the corresponding channel
    output $g(x_1,x_2)$ is an element of $\tilde{\mc{Y}}$. The existence
    of an induced matching of size $U$ corresponds to a scheme for
    computing the equality target function.}
    \label{Fig:Equality}
\end{figure}

Given an arbitrary MAC $\bm{G}$, create a bipartite graph $B$ as
follows. Let the vertices on the two sides of the bipartite graph
correspond to the $X$ different row and column indices of the channel
matrix $\bm{G}$. Fix an arbitrary subset $\tilde{\mc{Y}}$ of cardinality
$k$ of $\mc{Y}$. Place an edge in the bipartite graph $B$ between a node
$x_1$ on the left and a node $x_2$ on the right if
$g(x_1,x_2)\in\tilde{\mc{Y}}$, see Fig.~\ref{Fig:Equality}. 

An \emph{induced matching} $M$ in a bipartite graph $B$ is a subset of
edges such that i) no pair of edges in $M$ share a common endpoint and
ii) no pair of  edges in $M$ are joined by an edge in $B$.  Note that an
induced matching $M$ of size $U$ in $B$ corresponds to a zero-error
computation scheme for the equality function $\bm{A}_{=}$. This follows from the
observation that the induced matching provides a subset of channel
inputs $\{x_{1,1},x_{1,2},\ldots,x_{1,U}\} \subset \mc{X}_1$ and
$\{x_{2,1},x_{2,2},\ldots,x_{2,U}\} \subset \mc{X}_2$ such that the only
pairs of channel inputs for which the channel output is in $\tilde{\mc{Y}}$ are
given by $\{ (x_{1,k}, x_{2,k}): k \in \{1,2,\ldots,U\} \}$. The decoder
thus simply maps all  channel outputs in $\tilde{\mc{Y}}$ to $1$ and all
other channel outputs to $0$. 

A \emph{strong edge-coloring} of a graph $B$ is an edge-coloring in
which every color class is an induced matching, i.e., any two vertices
belonging to distinct edges with the same color are not adjacent. The
strong chromatic index $\chi_s(B)$  is the minimum number of colors in a
strong edge-coloring of $B$.  A simple argument in \cite{molloy97} shows
that for any graph $B$, 
\begin{equation*}
    \chi_s(B) \le 2\Delta^2(B),
\end{equation*}
where $\Delta(B)$ is the maximum degree of $B$. Thus, a graph $B$
contains an induced matching of size at least 
\begin{equation}
    \label{Eqn4a}
    \frac{m(B)}{\chi_s(B)} \geq \frac{m(B)}{2\Delta^2(B)}
\end{equation}
where $m(B)$ denotes the number of edges in $B$. 

Consider again the bipartite graph $B$ constructed from $\bm{G}$ for
some fixed subset $\tilde{\mc{Y}}$. From \eqref{Eqn4a} and the above
discussion, we see that $(\bm{A}_{=},\bm{G})$ is $0$-feasible if
\begin{equation}
    \label{Eqn4}
    \frac{m(B)}{2\Delta^2(B)} \geq U.
\end{equation}
We now show that this holds with high probability for a random channel
matrix $\bm{\msf{G}}$. Since we consider a random $\bm{\msf{G}}$, the
graph $\msf{B}$ is itself also random.

We start by deriving a lower bound on the number of edges $m(\msf{B})$
in $\msf{B}$. The event that a particular pair of vertices has an edge in
$\msf{B}$ is equivalent to the corresponding entry in the channel matrix
$\bm{\msf{G}}$ being an element of $\tilde{\mc{Y}}$, which happens with probability $k/Y$.
Since the $X^2$ entries of $\bm{\msf{G}}$ are independent, this implies
that the number of edges $m(\msf{B})$ is given by a binomial random
variable with mean $kX^2/Y$.  Thus, using the Chernoff
bound~\eqref{Eqn:Chernoff3}, that $X = 200U\ln(U)$ by assumption, and
that $Y/k < 32U$ by~\eqref{eq:k},
\begin{align}
    \label{eq:equality2}
    \Pp(m(\msf{B}) < kX^2/(2Y)) 
    &< \exp(- kX^2/(8Y)) \nonumber\\
    &< \exp( - U \ln^2(U)),
\end{align}
which converges to zero as $U\to\infty$.

We continue by deriving an upper bound on the maximum degree
$\Delta(\msf{B})$ of $\msf{B}$.  Let $\Delta_{L}(\msf{B})$ and
$\Delta_{R}(\msf{B})$ denote the maximum degree among the left-side and
right-side vertices, respectively. Note that $\Delta_{L}(\msf{B})$ and
$\Delta_{R}(\msf{B})$ are identically distributed, as the maximum value
among $X$ independent binomial random variables with mean $kX/Y$. Let
$\msf{z}$ be one such binomial random variable. Using the Chernoff bound
\eqref{Eqn:Chernoff2},
\begin{align*}
    \Pp(\Delta_{L}(\msf{B}) \leq 2kX/Y) 
    & = \Pp(\Delta_{R}(\msf{B}) \leq 2kX/Y) \\
    & = \bigl(\Pp(\msf{z} \leq 2kX/Y)\bigr)^X \\
    & \geq \bigl( 1 - (e / 4)^{kX/Y} \bigr)^X.
\end{align*} 
By the union bound, and using that $X=200U\ln(U)$ by assumption, that
$Y/k < 32U$ by~\eqref{eq:k}, and that $e/4 < \exp(-1/3)$, we  have 
\begin{align}
    \label{eq:equality3}
    \Pp(\Delta(\msf{B})> 2kX/Y ) 
    &= \Pp\bigl(\{\Delta_L(\msf{B})> 2kX/Y\} \cup \{ \Delta_R(\msf{B})> 2kX/Y\} \bigr) \nonumber\\
    & \leq  2\Bigl( 1 - \bigl( 1 - (e / 4)^{kX/Y} \bigr)^X \Bigr) \nonumber\\
    & \leq 2 X (e/4)^{kX/Y} \nonumber\\
    & \leq \exp\bigl( \ln(400U\ln(U)) - 200 U \ln(U)/(3\cdot32U) \bigr) \nonumber\\
    & = \exp\bigl( \ln(400\ln(U)) - 13\ln(U)/12 \bigr),
\end{align}
which converges to zero as $U\to\infty$. Using $Y/k \geq 16U$ by~\eqref{eq:k} and the union bound,
\begin{align*}
    \Pp\Bigl(\frac{m(\msf{B})}{2\Delta^2(\msf{B})} \geq U\Bigr) 
    &\geq \Pp\Bigl(\frac{m(\msf{B})}{2\Delta^2(\msf{B})} \geq \frac{Y}{16k}\Bigr) \\
    & \geq \Pp\Bigl(\bigl\{m(\msf{B})\geq kX^2 / (2Y)\bigr\} \cap
    \bigl\{\Delta(\msf{B}) \leq 2kX / Y \bigr\} \Bigr) \\
    &\geq 1 - \Bigl( \Pp\bigl(\bigl\{m(\msf{B})< kX^2 / (2Y)\bigr\}\bigr) + \Pp\bigl(\bigl\{ \Delta(\msf{B}) > 2kX / Y \bigr\} \bigr) \Bigr), 
\end{align*}
which, by \eqref{eq:equality2} and \eqref{eq:equality3}, converges to
one as $U\to\infty$. Combined with \eqref{Eqn4}, this shows that
\begin{equation*}
    \lim_{U\to\infty}\Pp\bigl((\bm{A}_{=},\bm{\msf{G}}) \text{ is $0$-feasible}\bigr)
    = 1,
\end{equation*}
thus proving the claim. \hfill\IEEEQED

\subsection{Proof of Theorem~\ref{Thm:converse_n}}
\label{sec:proofs_n}

Consider an arbitrary target function $\bm{A}$ and an arbitrary channel function $\bm{G}$. Recall the definition of a $\delta$-approximation function
in~\eqref{Eqn:ApproxFunction}. From Lemma~\ref{Lemma:Translation},
$(\bm{A},\bm{G}^{\otimes n})$ is $\delta$-feasible only if there exists
some $\delta$-approximation
$\bm{A}_\delta\in\mc{W}^{V_1\times V_2}$ of $\bm{A}$ such that $V_1, V_2 \le U$ and 
$(\bm{A}_\delta,\bm{G}^{\otimes n})$ is $0$-feasible. From the
construction in Lemma~\ref{Lemma:Translation}, we can assume without
loss of generality that $\mc{V}_1,\mc{V}_2\subseteq \mc{X}^n$. Furthermore,
we can assume without loss of generality that no two rows and no two
columns of $\bm{A}_\delta$ are identical. Hence,
$(\bm{A}_{\delta},\bm{G}^{\otimes n})$ is $0$-feasible only if there
exists a $V_1\times V_2$ ordered submatrix $\bm{S}$ of $\bm{G}^{\otimes
n}$ computing $\bm{A}_\delta$, as described in
Section~\ref{sec:proofs_preliminaries}.  In the following, denote by
$s\colon \mc{V}_1\times\mc{V}_2\to\mc{Y}^n$ the mapping corresponding to
$\bm{S}$. 

Consider now such a $V_1 \times V_2$ ordered submatrix $\bm{S}$ of
$\bm{G}^{\otimes n}$. For any $\mc{T} \subseteq \mc{V}_1 \times
\mc{V}_2$, let $s(\mc{T})\subseteq \mc{Y}^n$ denote the range of
$s(\cdot,\cdot)$, with the arguments restricted to the subset $\mc{T}$.
Let $\bm{\msf{v}}_1$ and $\bm{\msf{v}}_2$ be independent random
variables uniformly distributed over $\mc{V}_1$ and $\mc{V}_2$,
respectively. Consider the random vector
\begin{equation*}
    \bm{\msf{y}} =
    \begin{pmatrix}
        \msf{y}[1] & \cdots & \msf{y}[n]
    \end{pmatrix}
    \defeq s(\bm{\msf{v}}_1,\bm{\msf{v}}_2).
\end{equation*}
Then, for any $\bm{y}\in \mc{Y}^n$, we have 
\begin{equation}
    \label{Eqn:ProbDef}
    \Pp(\bm{\msf{y}} = \bm{y}) 
    = \frac{\big\lvert\bigl\{(\bm{v}_1, \bm{v}_2) \in \mc{V}_1\times\mc{V}_2: 
    s(\bm{v}_1, \bm{v}_2) = \bm{y} \bigr\}\big\rvert }{ V_1V_2 } \ ,
\end{equation}
and let $H_{\bm{S}}(\bm{\msf{y}})$ denote the corresponding entropy of
the random vector $\bm{\msf{y}}$.  The next result proves the existence
of a ``typical'' set.

\begin{lemma}
    \label{Lemma:typical}
    Let $\bm{S}$ be a $V_1\times V_2$-dimensional ordered submatrix of
    $\bm{G}^{\otimes n}$, and let $s\colon \mc{V}_1\times\mc{V}_2\to\mc{Y}^n$
    be the corresponding mapping. For any $\varepsilon > 0$, there exists
    a set $\mc{T}\subset \mc{V}_1\times\mc{V}_2$ such that
    \begin{subequations}
        \label{Eqn:typical}
        \begin{align}
            \label{Eqn:CardSet}
            \card{\mc{T}} & \geq \frac{\varepsilon}{1+\varepsilon} V_1V_2, \\
            \label{Eqn:DisColors}
            \card{s(\mc{T})} & \leq 2^{(1+\varepsilon)(2+ H_{\bm{S}}(\bm{\msf{y}})  )}.
        \end{align}
    \end{subequations}
\end{lemma}

The proof of Lemma~\ref{Lemma:typical} is presented in
Section~\ref{sec:proofs_n_typical}.  Consider now the event that
$(\bm{\msf{A}},\bm{G}^{\otimes n})$ is $\delta$-feasible for the random
target function $\bm{\msf{A}}$ (as introduced in
Example~\ref{eg:FnMAC}). As we have seen before, this implies the
existence of a $\delta$-approximation $\bm{\msf{A}}_\delta$ of
$\bm{\msf{A}}$ such that $(\bm{\msf{A}}_\delta,\bm{G}^{\otimes n})$ is
$0$-feasible.  Let $\bm{S}$ be the corresponding ordered submatrix of
$\bm{G}^{\otimes n}$ specifying the encoders, and let $\phi$ be the
corresponding decoder.  For fixed $\varepsilon>0$, let
$\mc{T}\subset\mc{V}_1\times\mc{V}_2$ be the typical set associated with
$\bm{S}$, as guaranteed by Lemma~\ref{Lemma:typical}.  Since $\phi$
correctly computes $\msf{a}_\delta(\cdot,\cdot)$ for all elements of
$\mc{V}_1 \times \mc{V}_2$, it does so in particular for all elements of
$\mc{T}$. More formally, 
\begin{equation*}
    \msf{a}_\delta(\bm{v}_1,\bm{v}_2)
    = \phi(s(\bm{v}_1,\bm{v}_2))
\end{equation*}
for all $(\bm{v}_1,\bm{v}_2)\in\mc{T}$.

Fix an ordered submatrix $\bm{S}$ of $\bm{G}^{\otimes n}$ of dimension
$V_1\times V_2$. Let $\mc{T}$ be the typical set
corresponding to $\bm{S}$.  Consider a random $\bm{\msf{A}}$, and let
$\mc{E}_{\bm{S}}$ be the event that there exists a
$\delta$-approximation $\bm{\msf{A}}_\delta$ of dimension $V_1\times
V_2$ and a mapping $\phi:s(\mc{T})\to\mc{W}$ such that
\begin{equation*}
    \msf{a}_\delta(\bm{v}_1,\bm{v}_2)
    = \phi(s(\bm{v}_1,\bm{v}_2)) 
\end{equation*}
for all $(\bm{v}_1,\bm{v}_2)\in\mc{T}$. From the discussion in the last
paragraph, we have
\begin{equation}
    \label{eq:mce}
    \Pp\bigl((\bm{\msf{A}},\bm{G}^{\otimes n}) \text{ is
    $\delta$-feasible}\bigr) \leq \Pp(\cup_{\bm{S}} \mc{E}_{\bm{S}}) \leq
    \sum_{\bm{S}}\Pp(\mc{E}_{\bm{S}}).
\end{equation}

We continue by upper bounding the probability of the event
$\mc{E}_{\bm{S}}$. Fix a mapping $\phi$ and let $\mc{A}_\delta^\phi$ denote the
set of distinct $V_1 \times V_2$ matrices $\bm{A}_\delta$ with entries
in $\mc{W}$ such that
\begin{equation*}
    a_\delta(\bm{v}_1, \bm{v}_2) 
    = \phi(s(\bm{v}_1, \bm{v}_2))
\end{equation*}
for all $(\bm{v}_1,\bm{v}_2)\in \mc{T}$. Using the union bound,  we have 
\begin{equation}
    \label{eq:mce2}
    \Pp(\mc{E}_{\bm{S}}) 
    \leq \sum_{\phi}\sum_{\bm{A}_\delta\in\mc{A}_\delta^\phi}
    \Pp(\text{$\bm{A}_\delta$ is a $\delta$-approximation of $\bm{\msf{A}}$}).
\end{equation}

The number of matrices in $\mc{A}_\delta^\phi$ is at most
\begin{align*}
    &\card{ \mc{A}_{\delta}^\phi } 
    \leq W^{V_1V_2 - \card{\mc{T}} }  
    \leq W^{V_1V_2/(1+\varepsilon)} 
    \leq W^{U^2/(1+\varepsilon)}  ,
\end{align*}
where the second inequality follows from \eqref{Eqn:CardSet} in
Lemma~\ref{Lemma:typical}.  Since there are at most
$W^{\card{s(\mc{T})}}$ mappings $\phi$ from $s(\mc{T})$ to $\mc{W}$, we
have
\begin{align}
    \label{Eqn:CardMatrices}
    \sum_\phi \card{ \mc{A}_{\delta}^\phi } \ \leq \ W^{\card{s(\mc{T})}} W^{U^2/(1+\varepsilon)} 
   \  \leq \ \exp\bigl( \ln(W)2^{(1+\varepsilon)( 2 +  H_{\bm{S}}(\bm{\msf{y}}))} 
    + \ln(W)U^2/(1+\varepsilon) \bigr)
\end{align}
where the last inequality follows from \eqref{Eqn:DisColors} in
Lemma~\ref{Lemma:typical}.

Consider then a fixed matrix $\bm{A}_\delta$. The next lemma upper bounds
the probability that this fixed $\bm{A}_\delta$ is, in fact, a
$\delta$-approximation of the random target function $\bm{\msf{A}}$.
\begin{lemma}
    \label{Lemma:ApproximationMatrix}
    Fix $0 < \delta < 1-1/W$ and an arbitrary matrix $\bm{A}_\delta$ of
    dimension $V_1\times V_2$ with $V_1,V_2\leq U$ and range of
    cardinality $W$. Let $\bm{\msf{A}}$ be the random target function of
    dimension $U\times U$ and range of cardinality $W$. Then
    \begin{equation*}
        \Pp(\text{$\bm{A}_\delta$ is a $\delta$-approximation of $\bm{\msf{A}}$})
        \leq \exp\bigl(2U\ln(U) - \alpha U^2\bigr) ,
    \end{equation*}
    with
    \begin{equation*}
        \alpha \defeq (1-\delta)\ln(W(1-\delta)) - (1-\delta).
    \end{equation*}
\end{lemma}

The proof of Lemma~\ref{Lemma:ApproximationMatrix} is presented in
Section~\ref{sec:proofs_n_approximation}.  Combining~\eqref{eq:mce2},
\eqref{Eqn:CardMatrices}, and Lemma~\ref{Lemma:ApproximationMatrix}
shows that for any $\bm{S}$,
\begin{equation*}
    \Pp(\mc{E}_{\bm{S}}) \leq \exp\Bigl( \ln(W)2^{(1+\varepsilon)( 2 +  H_{\bm{S}}(\bm{\msf{y}}))} 
    + 2U\ln(U) 
    - (\alpha-\ln(W)/(1+\varepsilon))U^2 \Bigr).
\end{equation*}

Substituting the above into \eqref{eq:mce}, we have
\begin{align}
    \label{eq:nupper}
    &\Pp\bigl((\bm{\msf{A}}, \bm{G}^{\otimes n}) \text{ is $\delta$-feasible}\bigr) \nonumber\\
    & \hspace{.18in}\leq \sum_{\bm{S}}\Pp(\mc{E}_{\bm{S}}) \nonumber\\
    & \hspace{.18in}\leq \sum_{\bm{S}} \exp\Bigl( \ln(W)2^{(1+\varepsilon)( 2 +  H_{\bm{S}}(\bm{\msf{y}}))} 
    + 2U\ln(U) 
    - (\alpha-\ln(W)/(1+\varepsilon))U^2 \Bigr) \nonumber\\
    &\hspace{.18in}\leq \exp\Bigl(2nU\ln(X) + \ln(W)2^{(1+\varepsilon)( 2 +  \max_{\bm{S}} H_{\bm{S}}(\bm{\msf{y}}))} 
    + 2(U + 1)\ln(U) 
    - (\alpha-\ln(W)/(1+\varepsilon))U^2 \Bigr) 
\end{align}
where the last inequality follows by noting that there are at most
$U^2X^{2nU}$ ordered submatrices $\bm{S}$ of $\bm{G}^{\otimes n}$ of dimension at most $U\times U$.

Now, set
\begin{equation*}
    \varepsilon \defeq \frac{1}{\tfrac{1}{2}\ln(W)-1},
\end{equation*}
and note that $\varepsilon\to 0$ as $U\to\infty$ since $W \geq
\omega(1)$ as $U\to\infty$. Recall that
\begin{equation*}
    \delta \leq 1/(2\ln(W))
\end{equation*}
by assumption. This implies that
\begin{align*}
    \alpha - \frac{\ln(W)}{1+\varepsilon}
    & = (1-\delta)\ln(W(1-\delta)) - (1-\delta) - \ln(W) + 2 \\
    & \geq (1-\delta)\ln(1-\delta) + \delta+1/2 \\
    & \geq 1/2.
\end{align*}
Hence, the right-hand side of \eqref{eq:nupper} converges to zero as
$U \to \infty$ if the following two conditions hold,
\begin{align*}
    n & \dotl U, \\
    \ln(W)2^{(1+ \varepsilon)(2+\max_{\bm{S}} H_{\bm{S}}(\bm{\msf{y}}) )} & \dotl U^2.
\end{align*}
In particular, since $W \leq U^2$ without loss of generality, and since
$\varepsilon \to 0$ as $U\to \infty$, this is the case whenever
\begin{align*}
    n &\dotl U, \\
    2^{\max_{\bm{S}} H_{\bm{S}}(\bm{\msf{y}})} &\dotl U^2 .
\end{align*}
Thus, if 
\begin{equation*}
    \lim_{U \to \infty} \Pp\bigl((\bm{\msf{A}}, \bm{G}^{\otimes n}) \text{ is $\delta$-feasible}\bigr) > 0 ,
\end{equation*}
then either 
\begin{subequations}
    \begin{equation}
        \label{Eqn:AltCondn}
        n \dotgeq U, 
    \end{equation}
    or there must exist a submatrix $\bm{S}$ of of $\bm{G}^{\otimes n}$ of
    dimension at most $U\times U$  such that 
    \begin{equation}
        \label{Eqn:Condn1}
        U^2 \dotleq 2^{H_{\bm{S}}(\bm{\msf{y}})}.
    \end{equation} 
\end{subequations}

Assume that the latter is true. Let $\bm{\msf{v}}_1, \bm{\msf{v}}_2$
denote independent random variables corresponding to the channel inputs
of the two users, as specified by the submatrix $\bm{S}$. Then we have 
\begin{align*}
    U^2 &\dotleq 2^{H_{\bm{S}}(\bm{\msf{y}})} \\
    &\leq  2^{H_{\bm{S}}(\bm{\msf{y}}, \bm{\msf{v}}_1)} \\
    &= 2^{H_{\bm{S}}(\bm{\msf{v}}_1) + H_{\bm{S}}(\bm{\msf{y}} | \bm{\msf{v}}_1) } \\
    &\leq 2^{\log U + H_{\bm{S}}(\bm{\msf{y}} | \bm{\msf{v}}_1) } ,
\end{align*}
which implies that
\begin{equation}
    \label{Eqn:Condn2}
    U \dotleq 2^{ H_{\bm{S}}(\bm{\msf{y}} | \bm{\msf{v}}_1) } .
\end{equation}
Similarly, we have 
\begin{equation}
    \label{Eqn:Condn3}
    U \dotleq 2^{ H_{\bm{S}}(\bm{\msf{y}} | \bm{\msf{v}}_2) } .
\end{equation}
From \eqref{Eqn:Condn1}, \eqref{Eqn:Condn2}, \eqref{Eqn:Condn3}, it
follows that there exists a joint distribution on $\mathcal{X}_1^n
\times \mathcal{X}_2^n \times \mathcal{Y}^n$ of the form 
\begin{equation*}
    p(\bm{v}_1, \bm{v}_2, \bm{y}) =  p(\bm{v}_1 ) \times  p(\bm{v}_2) \times \prod_{i=1}^{n} p(y[i] | v_1[i], v_2[i])
\end{equation*}
which satisfies 
\begin{align*}
    U^2 &\dotleq 2^{H(\bm{\msf{y}})} \leq 2^{\sum_{i=1}^{n} H(\msf{y}[i])} ,\nonumber\\
    U &\dotleq 2^{ H(\bm{\msf{y}} | \bm{\msf{v}}_1) } \leq 2^{\sum_{i=1}^{n} H(\msf{y}[i] | \msf{v}_1[i])} ,\nonumber\\
    U &\dotleq 2^{ H(\bm{\msf{y}} | \bm{\msf{v}}_2) } \leq 2^{\sum_{i=1}^{n} H(\msf{y}[i] | \msf{v}_2[i])}.
\end{align*}

We can then single-letterize the right-hand side of the above
inequalities in the usual way, see for example the proof of
\cite[Theorem~14.3.3]{cover91}. Then it follows that there exists a
joint distribution of the form $p(q)p(x_1 | q)p(x_2 | q) p(y | x_1,
x_2)$ such that  
\begin{align*}
    U^2 &\dotleq 2^{n(U)H\left(\msf{y} | \msf{q}\right)}, \\
    U &\dotleq 2^{n(U)H(\msf{y} | \msf{x}_1, \msf{q})}, \\
    U &\dotleq 2^{n(U)H(\msf{y} | \msf{x}_2, \msf{q})} .
\end{align*}
Thus, if \eqref{Eqn:Condn1} holds, then the above inequalities provide a
list of necessary conditions for $(\bm{\msf{A}}, \bm{G}^{\otimes n})$ to
be $\delta$-feasible with positive probability.  It follows easily that
the conditions above are also implied if the alternate condition in
\eqref{Eqn:AltCondn} holds, thus concluding the proof.  \hfill\IEEEQED

It remains to prove Lemmas~\ref{Lemma:typical} and
\ref{Lemma:ApproximationMatrix}.

\subsubsection{Proof of Lemma~\ref{Lemma:typical}}
\label{sec:proofs_n_typical}

Consider a variable-length binary Huffman code for the random variable
$\bm{\msf{y}}$ distributed according to~\eqref{Eqn:ProbDef}, and let
$\ell(\bm{\msf{y}})$ be the length of the codeword associated with
$\bm{\msf{y}}$. By \cite[Theorem~5.4.1]{cover91}, the expected length
\begin{equation*}
    L \defeq \E(\ell(\bm{\msf{y}})) 
\end{equation*}
of the code satisfies
\begin{equation}
    \label{Eqn:CodeLength}
    H_{\bm{S}}(\bm{\msf{y}}) 
    \leq L
    \leq H_{\bm{S}}(\bm{\msf{y}})+1.
\end{equation}

Let $\mc{C}\subset s(\mc{V}_1\times \mc{V}_2)$ denote the set of
$\bm{y}$ such that $\ell(\bm{y}) \leq (1+\varepsilon)L$ for some
$\varepsilon > 0$, and define
\begin{equation*}
    \mc{T} \defeq s^{-1}(\mc{C})
\end{equation*}
as the elements in $\mc{V}_1\times\mc{V}_2$ that are mapped into
$\mc{C}$. We have
\begin{align*}
     \card{s(\mc{T})}
     & = \card{\mc{C}} \\
     & \stackrel{(a)}{\leq} 2^{(1 + \varepsilon)L+1} \\
     & \stackrel{(b)}{\leq} 2^{(1 + \varepsilon)(H_{\bm{S}}(\bm{\msf{y}}) + 2)},
\end{align*}
where $(a)$ follows since there are at most $2^{(1+\varepsilon)L+1}$
binary sequences of length at most $(1+\varepsilon)L$ and each of them
can correspond to at most one value $\bm{y}\in\mc{C}$, and $(b)$ follows
from \eqref{Eqn:CodeLength}.  

On the other hand, we have 
\begin{align*}
    \card{\mc{T}} 
    & = \big\lvert \bigl\{(\bm{v}_1, \bm{v}_2) \in \mc{V}_1\times\mc{V}_2: 
    s(\bm{v}_1, \bm{v}_2) \in \mc{C} \bigr\} \big\rvert \\
    & \stackrel{(a)}{=} V_1V_2 \cdot \Pp(\bm{\msf{y}} \in \mc{C}) 
     = V_1V_2 \cdot \Pp\bigl( \ell(\bm{\msf{y}}) \leq (1 + \varepsilon)L \bigr) \\
    & \stackrel{(b)}{\geq} \frac{\varepsilon}{1+\varepsilon}V_1 V_2 , 
\end{align*} 
where $(a)$ follows from \eqref{Eqn:ProbDef} and $(b)$ follows from
Markov's inequality. Together, this shows the existence of a set $\mc{T}$
with the desired properties.\hfill\IEEEQED

\subsubsection{Proof of Lemma~\ref{Lemma:ApproximationMatrix}}
\label{sec:proofs_n_approximation}

Fix an arbitrary function $a_{\delta}\colon \mc{V}_1\times\mc{V}_2\to \mc{W}$
with $V_1,V_2 \leq U$, and fix arbitrary maps $f_1\colon \mc{U}_1 \to
\mc{V}_1$ and $f_2\colon \mc{U}_2 \to \mc{V}_2$. Denote by $\msf{z}_{u_1,
u_2}$ the indicator variable of the event
\begin{equation*}
    \big\{\msf{a}(u_1, u_2) = a_{\delta}(f_1(u_1), f_2(u_2))\bigr\}.
\end{equation*}
In words, $\msf{z}_{u_1, u_2}=1$ if the target function $\msf{a}(\cdot,
\cdot)$ is correctly approximated by $a_{\delta}(\cdot,\cdot)$ at
$(u_1,u_2)$. Since the entries of $\bm{\msf{A}}$ are uniformly
distributed over $\mc{W}$, we have
\begin{align*}
    \Pp(\msf{z}_{u_1, u_2} = 1) = 1/W . 
\end{align*}

Since the entries of $\bm{\msf{A}}$ are independent, the number of
message pairs for which the target function $\msf{a}(\cdot,\cdot)$ is
correctly computed using the approximation function
$a_{\delta}(\cdot,\cdot)$ is then described by a binomial random
variable 
\begin{equation*}
    \msf{z} 
    \defeq \sum_{u_1, u_2 \in \mc{U}} \msf{z}_{u_1, u_2} 
\end{equation*}
with mean $U^2/W$.  Thus the probability that for fixed maps $f_1,
f_2$, the function $a_{\delta}(\cdot,\cdot)$ is a
$\delta$-approximation of the random target function
$\msf{a}(\cdot,\cdot)$ is given by 
\begin{align*}
    \Pp(\msf{z} \geq (1-\delta) U^2 ) 
    & = \Pp\bigl( \msf{z} \geq (1 + W(1-\delta) - 1)U^2/W\bigr) \\
    & \stackrel{(a)}{\leq} \Bigl(\frac{\exp(W(1 - \delta) - 1)}
    {(W(1-\delta))^{W(1-\delta)}}\Bigr)^{U^2 / W} \\
    & \stackrel{(b)}{\leq} \exp(- \alpha U^2),
\end{align*}
where $(a)$ follows from the Chernoff bound \eqref{Eqn:Chernoff2}
and $(b)$ from the definition of $\alpha$.

The number of possible maps $f_1, f_2$ are at most $V_1^{U}$ and
$V_2^{U}$, respectively. By the union bound, the probability that
the function $a_{\delta}(\cdot,\cdot)$ is a $\delta$-approximation
of the random target function $\msf{a}(\cdot,\cdot)$ is then at most 
\begin{equation*}
    V_1^{U}V_2^{U}\exp(- \alpha U^2) 
    \leq \exp\bigl(2U\ln(U)- \alpha U^2\bigr),
\end{equation*}
thus concluding the proof.\hfill\IEEEQED

\appendices

\section{Proof of Lemma~\ref{Lemma:DistinctEntries} in Section~\ref{sec:proofs_identity}}
\label{sec:appendix_distinct}

We prove this result by posing it in the framework of the
coupon-collector problem, see, e.g., \cite[Chapter 3.6]{motwani95}. In
each round, a collector obtains a coupon uniformly at random from a
collection of $Y$ coupons. Let $\msf{z}$ denote the number of rounds
that are needed until the first time $X^2-X+1$ distinct coupons are
obtained. Then the event that $\msf{z}$ is at most $X^2$ is equivalent to the
number of distinct entries in the $X \times X$ random channel matrix $\bm{\msf{G}}$
being at least $X^2-X+1$. 

Let 
\begin{equation*}
    N \defeq X^2-X+1.
\end{equation*}
For the coupon collector problem, the minimum
number of rounds $\msf{z}$ needed to collect $N$ distinct coupons can be
written as 
\begin{equation}
\label{Eqn:Pis}
     \msf{z}  = \sum_{i=1}^{N} \msf{z}_i , \quad \msf{z}_i  \sim \Geom(p_i), \ p_i \defeq 1 - \frac{i - 1}{Y},
\end{equation}
where the $\msf{z}_i$'s are independent random variables and $\Geom(p_i)$
represents the geometric distribution with parameter $p_i$. Observe that
$p_1 \geq p_2 \geq \ldots \geq p_N$.

From the Chernoff bound \eqref{Eqn:Chernoff1},
\begin{align*}
    \Pp(\msf{z} > X^2) 
    & \leq \min_{t > 0} \exp(-tX^2) \prod_{i = 1}^{N} \E\bigl(\exp(t\msf{z}_i)\bigr) \\
    & \leq \min_{0 < t < -\ln(1-p_N)} \exp(-tX^2) \prod_{i = 1}^{N} \E\bigl(\exp(t\msf{z}_i)\bigr).
\end{align*}
We have
\begin{equation*}
    \E\bigl( \exp(t\msf{z}_i) \bigr) = \frac{p_i e^{t}}{1 - (1 - p_i)e^{t}}
\end{equation*}
for $t < - \ln( 1- p_i)$. Since the right-hand side is decreasing in
$p_i$, we have 
\begin{equation*}
    \E\bigl(\exp(t\msf{z}_i)\bigr) 
    \leq \E\bigl(\exp(t\msf{z}_{N})\bigr)
\end{equation*}
for every $i \in \{1,2,\ldots,N\}$. This implies that 
\begin{align}
    \label{eq:coupon1}
    \Pp(\msf{z} > X^2) 
    \leq \min_{0 <t < -\ln(1 - p_N)} \exp(-tX^2) 
    \Bigl( \frac{p_N e^{t}}{1 - (1 - p_N)e^{t}} \Bigr)^N.
\end{align}

Since $Y\geq e^3X^3$ by assumption, we have $-\ln( 1- p_N) > 2$ from \eqref{Eqn:Pis}.  Assume that this is the case in the following, and set $t
= 2$ in \eqref{eq:coupon1}. Then
\begin{equation} 
    \label{Eqn3}
    \Pp(\msf{z} > X^2) 
    \leq \exp\biggl(\!\! - \Bigl( 2X^2 - N \ln\Bigl( \frac{p_N e^2}{1 - (1 - p_N)e^2} \Bigr) \Bigr) \biggr).
\end{equation}
Now, since $Y \geq e^{3} X^3$, we have $p_N \geq 1-1/(e^3X)$, and thus
\begin{align} 
    \label{Eqn2}
    \frac{p_N }{1 - (1 - p_N)e^2} 
    & \ \leq \ \frac{ 1 - e^{-3} / X }{ 1 - e^{-1} / X }  \ \leq \ e^{ 1 / X }.
\end{align}
Here, the last inequality follows by setting $b = 1 / X$ in the
inequality 
\begin{align*}
    e^{b } - b e^{ b - 1} - 1 + b e^{ - 3}  \ge 0 \ \text{ for all $b \in [0, 1]$},
\end{align*}
which follows from the observation that the left-hand side evaluates to
zero at $b = 0$ and is monotonically increasing for $b \in [0, 1]$.

Substituting \eqref{Eqn2} into \eqref{Eqn3} and using the definition of
$N$ yields
\begin{align*}
    \Pp(\msf{z} > X^2) 
    & \leq \exp\Bigl( - \bigl( 2X^2 - N(2+1/X)\bigr) \Bigr) 
     \leq \exp\bigl(- (X-2) \bigr),
\end{align*}
thus concluding the proof. \hfill\IEEEQED

\section{Proof of Lemma~\ref{Lemma:DistinctEntries2} in Section~\ref{sec:proofs_balanced}}
\label{sec:appendix_distinct2}

Consider again the coupon collector problem as in
Appendix~\ref{sec:appendix_distinct}, and let $\msf{z}$ denote the
number of rounds required to collect $N$ distinct coupons. Then the
event that $\msf{z}$ is at least $\card{a^{-1}(\mc{W}_1)}$ is equivalent to
$\msf{N}_1$ being at most $N$.  Following the proof of
Lemma~\ref{Lemma:DistinctEntries}, we have from the Chernoff
bound~\eqref{Eqn:Chernoff1} that
\begin{align}
    \label{eq:coupon2}
    \Pp(\msf{N}_1 < N) 
    & = \Pp(\msf{z} > \card{a^{-1}(\mc{W}_1)}) \nonumber\\
    & \leq  \min_{0 <t < -\ln(1 - p_N)} \exp(-t\card{a^{-1}(\mc{W}_1)})
    \Bigl( \frac{p_N e^t}{1 - (1 - p_N)e^t} \Bigr)^N \nonumber\\ 
    & \leq \min_{0 <t < -\ln(1 - p_N)} \exp(-tcU^2)
    \Bigl( \frac{p_N e^t}{1 - (1 - p_N)e^t} \Bigr)^N,
\end{align}
where the last inequality follows since $\card{a^{-1}(\mc{W}_1)}\geq cU^2$ by
assumption, and with 
\begin{equation*}
    p_N \defeq 1-\frac{N-1}{Y}.
\end{equation*}

From the definition of $N$, we have $p_N > 2 / 3$ so that $-\ln(1-p_N) >
1$.  Choosing $t = 1 / 2$ in \eqref{eq:coupon2}, and noting that
\begin{equation*}
    \frac{p_Ne^{1/2}}{1 - (1 - p_N)e^{1/2}} 
    \leq \frac{2e^{1/2}/3}{ 1 - e^{1/2} / 3} 
    \leq e,
\end{equation*}
we obtain
\begin{align*}
    \Pp(\msf{N}_1 < N) 
     \leq \exp\bigl( - ( cU^{2}/2 - N)\bigr) 
     \leq \exp(- cU^{2}/6),
\end{align*}
thus proving the lemma.\hfill\IEEEQED

\end{document}